\documentclass[proceedings, preprint]{rmaa}

\usepackage{paralist}

\usepackage{psfrag,color}
\usepackage{url}
\usepackage{hyperref}
\usepackage[hyphenbreaks]{breakurl}

\SetYear{Revista Mexicana de Astronomía y Astrofísica}
\SetConfTitle{}

\title{Preliminary results from 5 years' spectral monitoring of Antares} 

\author{
  B. Oostra,\altaffilmark{1} 
  M.G. Batista,\altaffilmark{1}}

\altaffiltext{1}{Universidad de los Andes, Departamento de Física, Carrera 1 \#18a - 12, 111711, Bogotá, Colombia, (boostra@uniandes.edu.co), (mg.batistar@uniandes.edu.co).}

\shortauthor{Oostra \& Batista}
\shorttitle{Preliminary results from 5 years' spectral monitoring of Antares}

\listofauthors{B. Oostra, M.G. Batista}
\indexauthor{Oostra, B.}
\indexauthor{Batista, M.G.}

\abstract{We present preliminary results of 5 years’ monitoring of the radial velocity of Alpha Sco, performed at the Astronomical Observatory of the Universidad de Los Andes in Bogotá, Colombia. The data include 580 spectra acquired on 153 nights between March 2015 and March 2020. The aim of this study is to probe the dynamics of the star’s atmosphere on all possible time-scales through the variations in observed radial velocity. At present, our findings are consistent with previous results from other observers, and the combination of older and new data make it possible to assess the several periodicities. A detailed study of these results, including the convective motions in the photosphere, is still in progress.}

\resumen{Presentamos resultados preliminares de 5 años de monitoreo de la velocidad radial de Alpha Sco, realizado en el Observatorio Astronómico de la Universidad de Los Andes en Bogotá, Colombia. Los datos incluyen 580 espectros adquiridos en 153 noches entre marzo de 2015 y marzo de 2020. El objetivo de este estudio es investigar la dinámica de la atmósfera de la estrella en todas las escalas de tiempo posibles a través de las variaciones en la velocidad radial observada. En la actualidad, nuestros hallazgos son consistentes con resultados previos de otros observadores, y la combinación de datos antiguos y nuevos hace posible evaluar las diversas periodicidades. Un estudio detallado de estos resultados, incluyendo los movimientos convectivos en la fotósfera, aún está en progreso.}

\addkeyword{Spectroscopy}
\addkeyword{Red Supergiant}
\addkeyword{Radial Velocity}
\addkeyword{Stellar Atmospheres}

\begin{document}
\maketitle

\section{Introduction}
\label{sec:intro}

More than a century ago Wright (1906) (at Lick Observatory) noted the variability of Antares’ radial velocity (RV). Further studies were conducted by Halm (1909), Lunt (1916), Spencer-Jones (1928), Evans (1961), Smith et al. (1989) and Pugh \& Gray (2013A).

These studies reveal a six-year oscillation with a semi-amplitude of about 3 km/s around a mean velocity of -3 km/s. Additional to the 6-year period, several observers have noted the existence of faster oscillations with a period of a few months. Other Red Supergiant (RSG) stars such as Betelgeuse and Alpha Herculis exhibit similar behaviors \citep{Sm:1989}.

The importance of these observations lies in the fact that the cause or excitation mechanism of RV oscillations in RSG is still not fully understood. Early observers considered the possibility of binary-star orbital motion (i.e. Halm (1909), Lunt (1916)), but it has already become clear that we are seeing single-star radial pulsations (i.e. Evans (1961),  Smith et al. (1989)). Spencer-Jones (1928) proposed that the variations were $\delta$ Cephei like; however, RSG are definitely outside the instability strip on the HR diagram.

There seems to be no report of a series of observations of Antares covering densely and consistently a complete 6-year period with the same instrumentation. This is what we attempted to do, taking advantage of our equatorial location which gives us longer observing seasons than previous observers at more northern latitudes. Additional goals included: improve the time-coverage and cadence of previous work; combine new observations with published results to construct a longer time baseline; compare new and old data to detect eventual changes; take account of convection effects; search for variability in the granulation pattern; and search for fast (day- or hour-scale) variations.

We could not reach the 6-year goal due to the closing of our campus from March 2020 till end 2021. New observations began in September 2023 using an improved spectrograph. Neither have we completed the study of convection effects: the velocities presented in this initial report are simple averages of 15 selected spectral lines (Table \ref{tab:lines}). A simplistic approach would assume that the effect of convection would just introduce an additive constant; but this is not the case, because we observed the granulation pattern to be variable, a fact also reported by Gray and Pugh (2012).

In the next section we present details of the instruments and observations; in Section 3 we summarize our results, compare with previously published data and highlight some notable findings; and Section 4 gives some plans and expectations of future work.

\section{Observations}
\label{sec:observation}

We secured a total of 580 spectra on 153 nights from March 2015 to March 2020 (Table \ref{tab:nights}). 

\begin{table}[!t]\centering
  \setlength{\tabnotewidth}{\columnwidth}
  \tablecols{3}
  \caption{Summary of observations} \label{tab:nights}
  \begin{tabular}{lcc}
    \toprule
    Year & \multicolumn{1}{c}{Nights} & \multicolumn{1}{c}{Spectra} \\
    \midrule
2015 & 27     & 53      \\ 
2016 & 33     & 204     \\
2017 & 13     & 49      \\
2018 & 35     & 142     \\ 
2019 & 37     & 106     \\ 
2020 & 8      & 26      \\ 
    \bottomrule
  \end{tabular}
\end{table}

From our location in Bogotá, Colombia, at 4.6° northern latitude, Antares can be studied, in principle, from mid-January to the first days of November, the limits being imposed by the mountains to the East and the city buildings to the West of the observatory. Observing dates were not planned systematically; they are the result of the star's visibility, the observers' available time, and the weather. Additionally, observations were impossible during the best part of 2017 because the camera failed and a new one had to be purchased and installed.

Some details of the instrumentation:
\begin{itemize}
    \item Telescope: Meade LX200 (40 cm)
    \item Spectrograph: Espartaco \citep{Oos:2011} 
    \item Spectral resolution: 31000
    \item Calibration: Hollow-cathode Th-Ar lamp on separate exposures, and a Neon spectrum on all exposures.
    \item Exposure time: 20 minutes
    \item S/N ratio: $\sim$ 100
    \item Uncertainty: $\sim$ 0.1 km/s judged from the dispersion of the results.
\end{itemize}

The wavelength range was selected between 6140 and 6265 \AA~to avoid strong telluric lines and include strong Th-Ar and Neon lines (three Neon lines, at 6143, 6164 and 6217 Angstrom, were present in all spectra through a second fiber and made it possible to monitor the instrument's shifts between the stellar and calibration spectra). Unfortunately, this spectral range coincides with a strong TiO band which affects line depth measurements.

\begin{table}[!t]\centering
  \setlength{\tabnotewidth}{0.8\columnwidth}
  \tablecols{4}
  \caption{Spectral lines used for velocity measurements \tabnotemark{a}} \label{tab:lines}
  \begin{tabular}{cccc}
    \toprule
    Line & \multicolumn{1}{c}{$\lambda$} & \multicolumn{1}{c}{Line} & \multicolumn{1}{c}{$\lambda$}\\
    \midrule
Zr I & 6143.2     & Sc I & 6210.658  \\
Fe I & 6151.6169  & V I  & 6216.3643 \\
Na I & 6154.2255  & Fe I & 6219.2801 \\
Fe I & 6157.7275  & V I  & 6251.8231 \\
Ca I & 6162.173 & Fe I & 6252.5546 \\
Fe I & 6173.3339  & Fe I & 6254.2573 \\
Fe I & 6180.2018  & Ti I & 6261.0988 \\
Fe I & 6200.3121  &      &          \\
    \bottomrule
    \tabnotetext{a}{The wavelengths are from the VALD server \citep{Ry:2015}}
  \end{tabular}
\end{table}

For the conversion from the observatory reference to the heliocentric reference we used the tool included in the ISIS software (ISIS, 1997).

\section{Results and discussion}

\begin{figure}[!t]
  \includegraphics[width=\columnwidth]{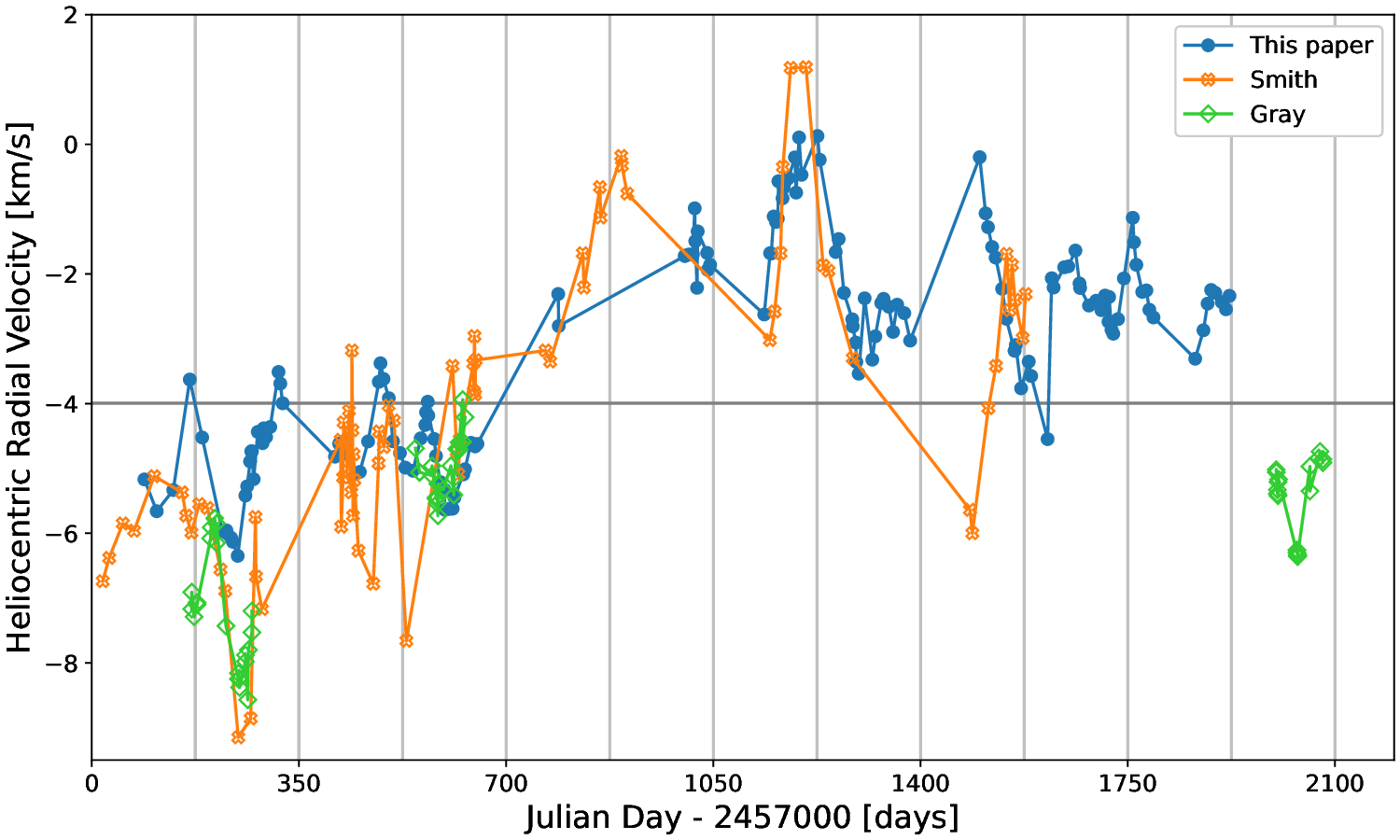}
  \caption{Nightly averages of our measured velocities (blue dots), compared with results from Smith et al. (1989) (orange dots) and Pugh \& Gray (2013) (green dots). The uncertainty of a typical average is about 0.1 km/s. Vertical lines mark steps of 175 days, where the secondary peak of the combined periodogram was found. }
  \label{fig:espectro}
\end{figure}

Figure ~\ref{fig:espectro} shows our results in blue dots; the numerical data are available in machine-readable format\footnote{Observational data are available in the repository: \url{https://github.com/MGBatista1/Antares_Uniandes.git}}. Each dot represents the average of the velocities from one night, weighted by the amount of signal present in each individual spectrum. The mean heliocentric velocity of our measurements is -3.14 km/s, with an amplitude $\pm$ 3 km/s. This result is consistent with previous reports from Smith et al. (1989) and Pugh \& Gray (2013A). We also combined these three studies in order to find a global oscillation behavior and the mean heliocentric velocity obtained is -3.99 km/s, where several smaller oscillations with shorter periods are also evident.


Superimposed on our results are Smith's and Gray's data, shifted forward by time-intervals determined using a cross-correlation method in order to ensure the best fit. For this comparison we shifted Smith's data forward by 10946 days; assuming 5 cycles to have elapsed between Smith's and our observations, this gives a mean period of 2189 days. Gray’s first-season data are advanced two periods, and his other two seasons only one period.

We calibrated the spectra directly with calibration lamps, and the same was done by Pugh and Gray (2013A); however, Smith et al (1989) did it by comparing some metal lines of Antares with the same lines in a spectrum of Arcturus recorded on the same night, and using a known or reported value of Arcturus' velocity. The difference in methods may introduce a small discrepancy in the radial velocity zero point. For now, we have not considered this detail.

We observe that some short-period oscillations are similar in the three mentioned datasets. The highest peak observed by Smith et al.~(1989) (plotted here at Day $\sim$1200) was also reported by Wright (1906), but appears lower and broader in our data. Around Day 500 we see an 80-day periodic variation which resembles the 100-day oscillation reported by Pugh and Gray (2013B).

\begin{figure}[!t]
  \includegraphics[width=\columnwidth]{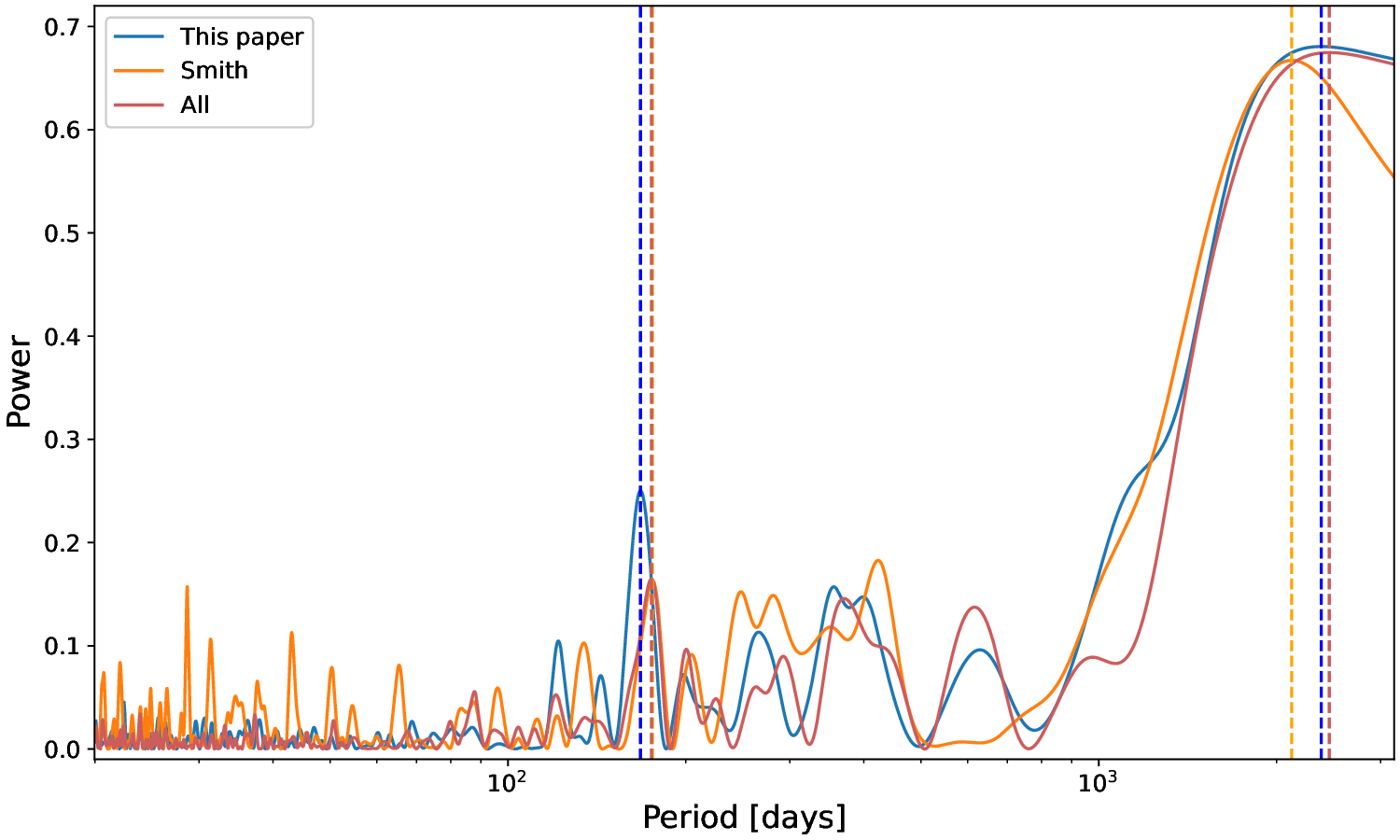}
  \caption{Lomb-Scargle periodogram of our data (blue line), Smith data (orange) and combined data (light red). Vertical lines show periods close to the two highest powers of our data.}
  \label{fig:periodograma}
\end{figure}

As to the six-year oscillation, Figure~\ref{fig:periodograma} shows a Lomb-Scargle periodogram of our data, including a broad peak around 2380 $\pm$ 11 days. A periodogram of Smith's results is similar; it peaks at 2120.5 $\pm$ 1.7 days. The three datasets combined give us a narrow peak at 2455.8 $\pm$ 7.4 days. Errors are estimated by applying a Monte Carlo method. For comparison, Pugh and Gray (2013A) combined their data with all available previous results and report a period of 2167 $\pm$ 5 days. The exact value of the six-year period needs not be unique nor constant.

Our periodogram (Fig.~\ref{fig:periodograma}) also shows a conspi-cuous secondary peak at 167.6 $\pm$ 0.08 days. Smith's and the combined data shows a peak at 175.6 $\pm$ 0.16 days and 174.9 $\pm$ 0.05 days, respectively, which are close to the theoretical value of 180 days computed by Stothers, R. (1969) for the fundamental mode radial pulsation on Antares. Likewise, this is close to the double of the 80-day period mentioned above; but the graph shows no sign at 80 days, nor at 100 days as given by Pugh and Gray (2013B). Vertical lines in Figure 1 show steps of 175 days, and several peaks tend to occur close to them.

In Figure~\ref{fig:espectro}, around Day 200 where the minimum radial velocity occurs, Smith's and Gray's results show a larger negative velocity than ours; between Day 500 and 600 all results seem to match; and around Day 1200, where the maximum RV occurs, Smith's measurements are higher than ours. This suggests the amplitude has decreased during the last years. Smith et al.~(1989) has commented on the varying amplitude of Alpha Ori. More observations will help to confirm and characterize such behavior. 

A decreasing amplitude is to be expected when no valve mechanism is available to pump energy into the pulsation as occurs in Cepheids. This suggests a damped oscillation, excited irregularly by sporadic or stochastic events. Our observations around Days 1500 and 1600 might have witnessed such an event, because after the maximum, instead of declining towards Gray’s dataset, the RV rises sharply and remains above the expected behavior. Smith et al.~(1989) also reports a ``missing half-cycle'' which may be related to this non-periodical driving force. The resulting phase shifts will be a limitation when comparing or combining several datasets.

Some of our datasets acquired during single nights seem to indicate significant variations of the radial velocity ($\sim$ 0.5 km/s) in about an hour. Our 20-minute integration time didn't allow us to confirm this, but presently we are using a faster spectrograph (Oostra \& Batista in prep), allowing 5-minute exposures, to gather more evidence on this aspect.

\begin{figure}[!t]
  \includegraphics[width=\columnwidth]{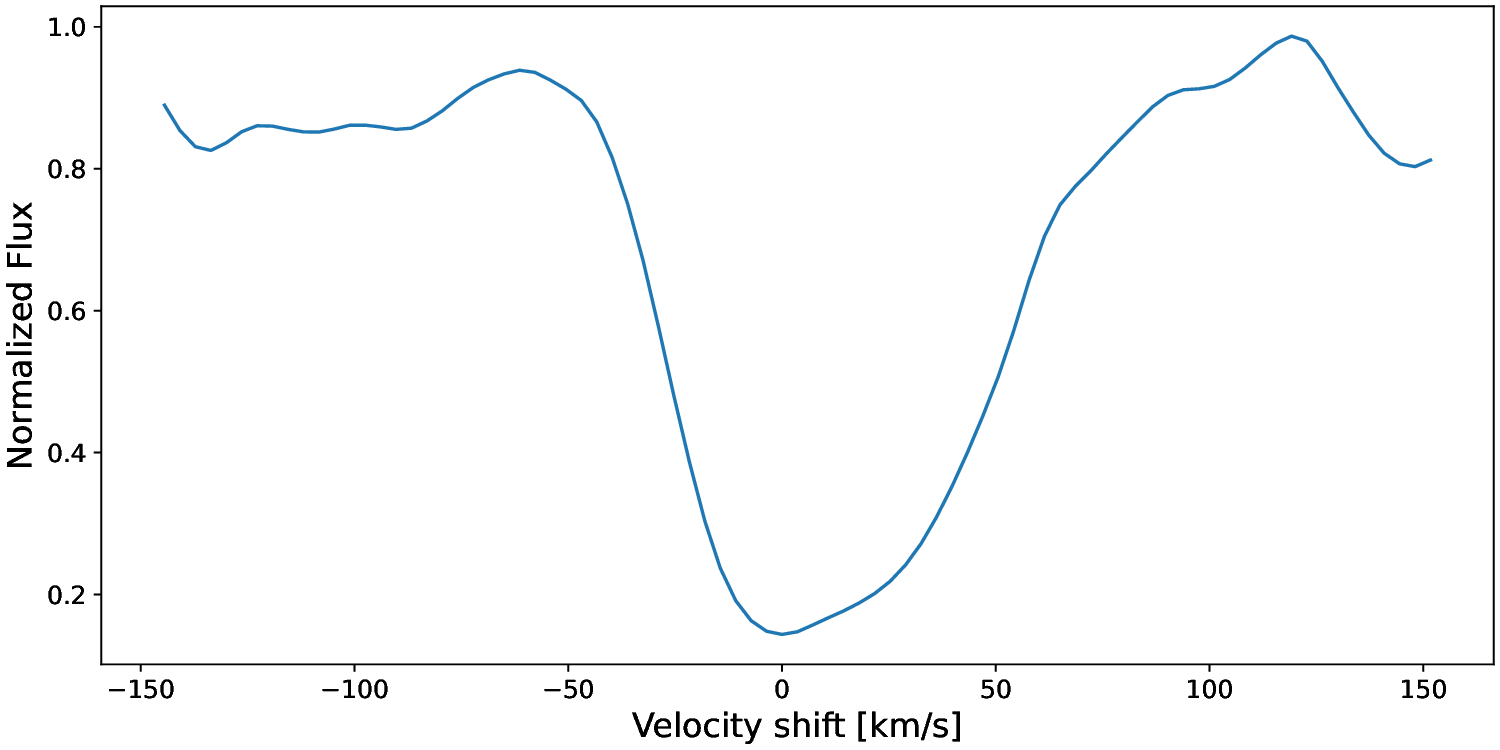}
  \caption{Profile of the H-Alpha absorption line in Antares. The asymmetrical shape does not change appreciably between observations.}
  \label{fig:Halpha}
\end{figure}

Smith et al.~(1989) showed that the core of the H-Alpha line gives a velocity up to 15 km/s more negative than metallic lines; they report a large scatter of this excess, but also a systematic variation which might be in phase with the 6-year oscillation. Moreover, the hydrogen line is quite asymmetrical; the barycenter of the whole profile gives a velocity similar to the metallic lines, but the deepest core is notably shifted blueward; this may be related with a H-rich stellar wind. In 2018 we took some additional spectra around H-alpha to check these facts; we found the hydrogen line profile evidently asymmetrical (Figure~\ref{fig:Halpha}) and the inner core shifted blueward by some 12 km/s (Figure 4). These details warrant more investigation.

\begin{figure}[!t]
  \includegraphics[width=\columnwidth]{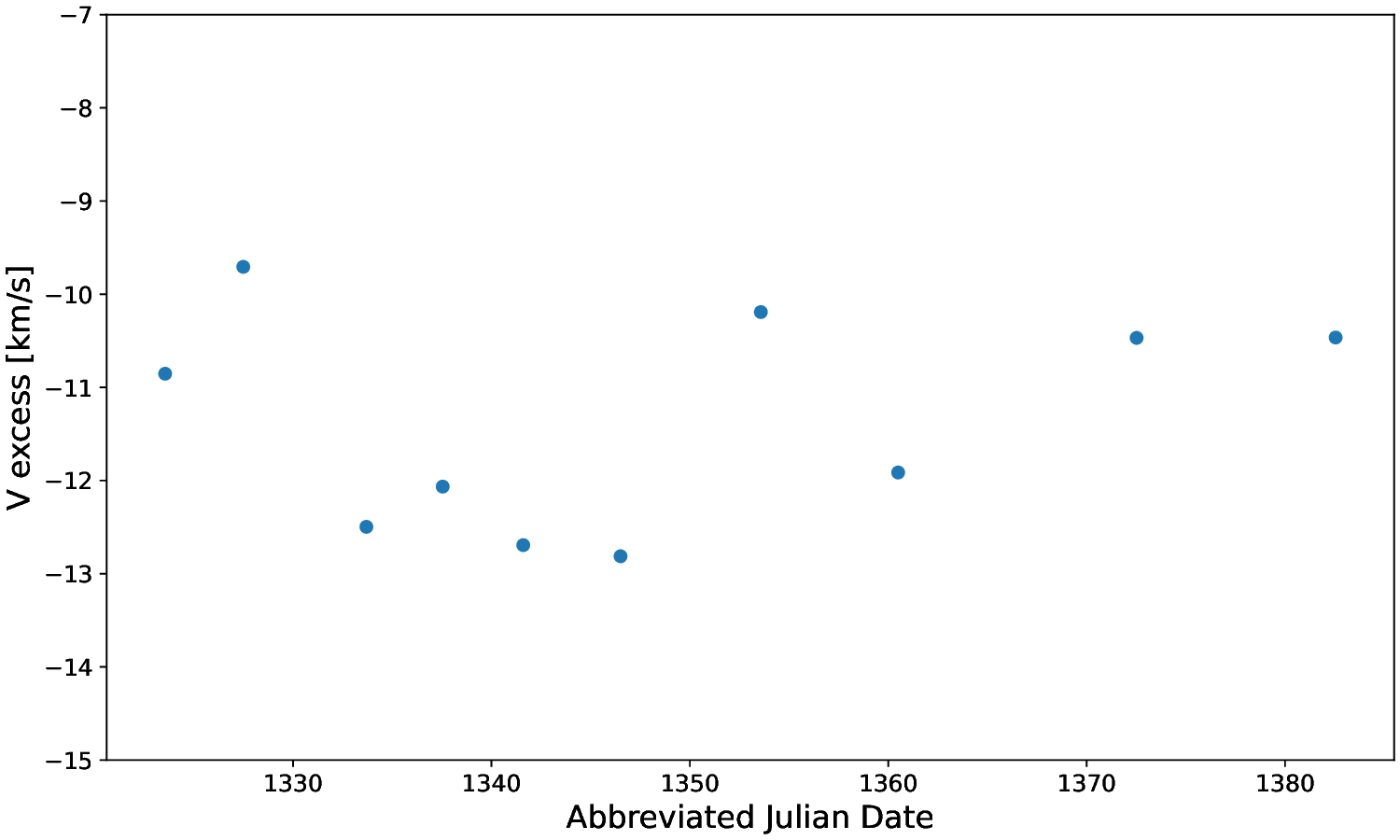}
  \caption{Velocity excess of the deepest core of the H-Alpha line, respect to the metallic lines, measured during August and September 2018.}
  \label{fig:Vexc}
\end{figure}

\section{Conclusions and Future work}
As to our observatory we may conclude that, despite the adverse environmental conditions, it has proven possible to obtain a meaningful set of observed data on stellar radial velocities.

Our findings are consistent with previous results from other observers, including a 6-year main pulsation and several shorter periods, the most prominent being an oscillation of roughly 175 days, around a mean velocity of -4 km/s.

A comparison of our results with previous observations hints at a variable amplitude.

For the near future we are planning to acquire new data and compare them with our previous results, in order to conclude more about possible changes in the amplitude of the oscillations. Also, our new spectrograph enables us to take 10 spectra per hour, which makes it possible to confirm or reject the fast variations. As our observations span a greater time interval, more will be known about the several oscillation periods.

A better understanding of the (changing) granulation pattern will give us better estimates of the velocity of the outer layers of the star, and, by the way, also for some deeper layers. Also, as has been shown elsewhere \citep{Oos:2022}, due to convection effects it is convenient to limit the analysis to a single chemical species; preferably neutral iron \citep{D:1981}. These facts justify a complete reevaluation of the measured velocities. A study along this line has been initiated by Luisa Rodríguez (2018) using the first three years of observation.

The Hydrogen Alpha line profile and differential blueshift will also demand more attention. Finally, comparisons with other similar stars will be useful.

\section*{Acknowledgements}

This work has made use of the VALD database, operated at Uppsala University, the Institute of Astronomy RAS in Moscow, and the University of Vienna.

\end{document}